\documentclass{article}
\usepackage{spconf,amsmath,graphicx}
\usepackage{xcolor,array}

\newcommand\blfootnote[1]{%
	\begingroup
	\renewcommand\thefootnote{}\footnote{#1}%
	\addtocounter{footnote}{-1}%
	\endgroup
}

\title{Deep Residual Echo Suppression and Noise Reduction:\\ A Multi-Input FCRN Approach in a Hybrid Speech Enhancement System}
\name{Jan Franzen and Tim Fingscheidt}
\address{Institute for Communications Technology, Technische Universit\"at Braunschweig,\\ Schleinitzstr. 22, 38106 Braunschweig, Germany\\
	{\small \tt \{j.franzen, t.fingscheidt\}@tu-bs.de}}

\begin{document}
\ninept
\maketitle
\begin{abstract}
	Deep neural network (DNN)-based approaches to acoustic echo cancellation (AEC) and hybrid speech enhancement systems have gained increasing attention recently, introducing significant performance improvements to this research field. 
	Using the fully convolutional recurrent network (FCRN) architecture that is among state of the art topologies for noise reduction, we present a novel deep residual echo suppression and noise reduction with up to four input signals as part of a hybrid speech enhancement system with a linear frequency domain adaptive Kalman filter AEC. In an extensive ablation study, we reveal trade-offs with regard to echo suppression, noise reduction, and near-end speech quality, and provide surprising insights to the choice of the FCRN inputs: In contrast to often seen input combinations for this task, we propose not to use the loudspeaker reference signal, but the enhanced signal after AEC, the microphone signal, and the echo estimate, yielding improvements over previous approaches by more than $0.2$\,PESQ points.
\end{abstract}

\begin{keywords}
	speech enhancement, residual echo suppression, noise reduction, fully convolutional recurrent network
	\blfootnote{\scriptsize$\copyright$ 2022 IEEE. Personal use of this material is permitted. Permission from IEEE must be obtained for all other uses, in any current or future media, including reprinting/republishing this material for advertising or promotional purposes, creating new collective works, for resale or redistribution to servers or lists, or reuse of any copyrighted component of this work in other works.}
\end{keywords}

\section{Introduction}
\label{sec:intro}

Speech enhancement systems are deployed in various applications such as smartphones, home and web conference devices, automotive hands-free systems and many more. 
An often used arrangement of the task is twofold: as a first stage, the undesired echo component stemming from the system's own loudspeakers should be removed from the microphone signal, which is referred to as acoustic echo cancellation (AEC) and is typically performed with a linear adaptive filter.
As a second stage, these systems should further improve the signal by performing residual echo suppression (RES) and noise reduction (NR)---usually done simultaneously---to obtain an ideally echo- and noise-free near-end speech component.

In this typical two-stage arrangement for speech enhancement, the application of a DNN as second stage (RES and NR) gained increasing attention with early investigations of feed-forward networks~\cite{Schwarz_NN_FF_RES, carbajal_RES_ICASSP}, convolutional networks bringing further improvements more recently~\cite{Seidel_ySquare_arxiv, Halimeh_RES_ICASSP}, some even being fully synergistic with the first stage~\cite{Haubner_SynergisticKalmanRES_arxiv}, and many more.  
In the meantime, also fully learned deep AEC approaches were proposed, where a single network incorporates the tasks of AEC, RES, and NR, e.g., \cite{wang_NN_AEC_19, fazel_DeepMultitaskAEC} or further investigated in~\cite{Franzen_AEC_NetShell_ICASSP}. Although showing an impressive suppression performance, however, these are often accompanied by some near-end speech degradation and---at least for now---hybrid approaches are still one step ahead as can be seen with the leading model of the AEC Challenge on ICASSP 2021 being a hybrid approach~\cite{amazon_PercepNet}. 

Although many proposals for hybrid approaches can be found in the literature recently, we find that a solid investigation with regard to the inputs of the second stage (RES/NR) network is lacking: While in~\cite{carbajal_RES_ICASSP} only the impact of very few input combinations on a rather simple feed-forward neural network is investigated without explicit consideration of noise reduction capabilities, we observe that for convolutional neural networks the utilized inputs to the second stage network are often just stated without further questioning, reasoning, or verification. 

Just recently, the fully convolutional recurrent network (FCRN) topology has proven itself very successful for noise reduction, as for example found in the works of Strake et al.~\cite{strake_SingleStage_ICASSP, strake_DNSchallenge_INTERSPEECH} or Xu et al.~\cite{Xu_PESQNet_arxiv}.
Accordingly, in this work we investigate the FCRN topology as second stage for RES and NR \textit{in the context of a hybrid speech enhancement system} with a frequency domain adaptive Kalman filter for AEC~\cite{enzner_vary_fdaf, jung_elshamy_SAEC}. Our contributions are as follows: 
First, we propose a multi-input FCRN for RES/NR and investigate all possible input signal combinations, revealing trade-offs between different performance aspects and providing surprising insights to simple design choices. 
Second, we investigate whether the improvements of the magnitude-bounded complex mask-based procedure in~\cite{strake_DNSchallenge_INTERSPEECH, Seidel_ySquare_arxiv} can be transferred to this task as well. Third, with the right choices, we improve performance over state of the art.
It should be noted that the findings of this work are not at all requiring a Kalman filter AEC, but are easily transferable to other hybrid speech enhancement systems and network topologies.

The remainder of this paper is structured as follows: In Section~\ref{sec:system}, a system overview of the hybrid speech enhancement system including the framework and network topology is given. Experimental variants for our investigation are described in Section~\ref{sec:variants}.
In Section~\ref{sec:results}, the experimental validation and discussion of all approaches is given. Section~\ref{sec:conc} provides conclusions.

\section{System Model and Network Topology}
\label{sec:system}

\begin{figure*}[htb!]
	\centering
	\includegraphics[width=\textwidth]{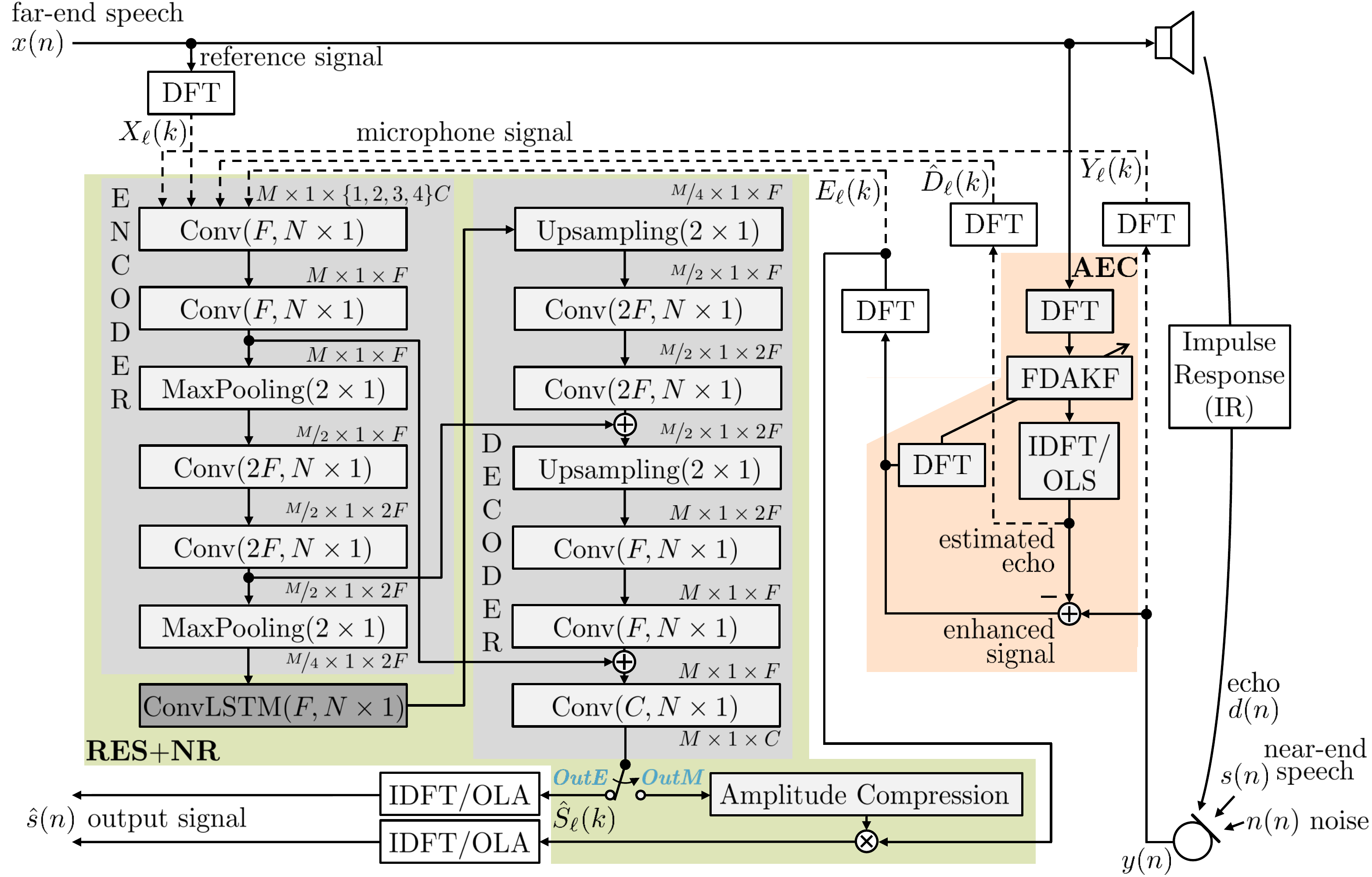}
	\vspace{0.5mm}
	\caption{\textbf{System model} and \textbf{FCRN topology for RES and NR} with options for input combinations (dashed lines) and output types (switch).}
	\label{fig:network}
\end{figure*}

For this work, a single-channel hands-free speech enhancement scenario is considered. The acoustic setup is outlined in Figure~\ref{fig:network} and shall cover typical single- and double-talk scenarios.
Following the procedure described in~\cite{Franzen_AEC_NetShell_ICASSP, wang_NN_AEC_19}, the TIMIT dataset~\cite{TIMIT} is used to set up far-end speech~$x(n)$ and near-end speech~$s(n)$ at a sampling rate of~\hbox{$16$\,kHz}.
At the microphone, near-end speech~$s(n)$ is superimposed with background noise~$n(n)$ and echo signals~$d(n)$. Echoes~$d(n)$ are generated by imposing loudspeaker nonlinearities~\cite{wang_NN_AEC_19} on far-end signals~$x(n)$ and applying impulse responses (IRs) created with the image method~\cite{image_method}. As in related work~\cite{Franzen_AEC_NetShell_ICASSP}, IRs are set to a length of $512$~samples with reverberation times $T_{60} \! \in \! \{0.2, 0.3, 0.4\}$\,s for training and validation, and $0.2$\,s for test mixtures. Background noise~$n(n)$ from the QUT dataset~\cite{QUT} is used for training and validation. For the test set unseen additive noises are used: babble, white noise, and operations room noise from the NOISEX-92 dataset~\cite{noisex}. 

With signal-to-echo ratios (SER) being randomly selected from $\{-6, -3, 0, 3, 6, \infty\}$\,dB per mixture and signal-to-noise ratios (SNR) from $\{8, 10, 12, 14, \infty\}$\,dB per mixture, this leads to a total of 3000 training and 500 evaluation mixtures. 
With \emph{unseen} speakers and utterances from the CSTR VCTK database~\cite{VCTK} as well as \emph{unseen} impulse responses and noise sequences, the test set consists of 280 mixtures where 
SER and SNR are set to $3.5$\,dB and $10$\,dB, respectively. 
As in our previous work, we additionally provide an evaluation with the test files consisting of echo only, near-end noise only or near-end speech only, thereby revealing deeper insights into the different network performance aspects.

\subsection{First Stage: Kalman-Based AEC}

As first stage a traditional AEC algorithm, namely the well-known frequency domain adaptive Kalman filter (FDAKF)~\cite{enzner_vary_fdaf}, is deployed. It is based on overlap-save processing and can therefore be easily synchronized with the second stage's network by using the same frame shift (here: $R \! = \! 256$). It is set to operate at a sampling frequency of~\hbox{$16$\,kHz} with a DFT size~\hbox{$K_{\textrm{AEC}} \! = \! 1024$}. Further parameters are: forgetting factor~\hbox{$A \! = \! 0.998$}, overestimation factor~\hbox{$\lambda \! = \! 1.5$}, and measurement noise covariance matrix $\mathbf{\Psi}_{ss}$~smoothing factor~\hbox{$\beta \! = \! 0.5$}. For the sake of brevity, the reader is referred to \cite{jung_elshamy_SAEC} or \cite{Franzen_Book_ICC} for a detailed description of the FDAKF. In a traditional system, the FDAKF would usually be combined with a closely tied RES approach~\cite{KuechEnzner_StateSpacePartitionedFDAF, franzen_RES_ICASSP}. In this work, however, the FCRN as second stage for RES and NR shall be investigated.
Since we are not exclusively bound to the Kalman filter (unlike for example~\cite{Haubner_SynergisticKalmanRES_arxiv}, where a synergistic approach is proposed), the first stage could certainly be substituted by any other AEC algorithm that provides the respective signals as used by the second stage.

\subsection{Second Stage: FCRN-Based RES and NR}

Our proposed system uses the well-performing fully convolutional recurrent network (FCRN)~\cite{strake_SingleStage_ICASSP, strake_DNSchallenge_INTERSPEECH} as second stage. It is depicted in the green box of Figure~\ref{fig:network} and operates on discrete Fourier transform (DFT) inputs, e.g., $Y_\ell(k)$, with frame index $\ell$ and frequency bin $k$.
It follows a convolutional encoder-decoder structure where the parameters of the layers are Conv(\textit{\# filters, kernel dimensions}) with Leaky ReLU activations~\cite{leakyReLU} and linear activations for the output layer, MaxPooling(\textit{pool dimensions}) over the feature axis, and likewise for upsampling. Two skip connections are employed. The final output is of dimension~$M \! \times \! 1 \! \times \! C$.
To enable the network to model temporal context, a convolutional LSTM~\cite{ConvLSTM} with $F \! = \! 88$~filter kernels of size~$N\!\times\! 1$ is placed at the bottleneck, with~$N \! = \! 24$.
Denoted as \textit{feature axis $\times$ time axis $\times$ number of feature maps}, the feature dimensions can be seen at the in- and output of each layer. As the network is subsequently processing single input frames, the time axis value is set to~$1$. 

Input features and training targets are extracted with a frame length of $K \! = \! 512$~samples and aforementioned frame shift of~$R \! = \! 256$~samples. A square root Hann window is applied and complex spectra are obtained by computing the~$512$-point DFT. They are separated into real and imaginary parts which leads to $C\!=\!2$ channels, and zero-padded to feature maps of height $M\!=\!260$.

As we investigate the FCRN's performance for RES and NR, clean speech is selected as training target and the MSE loss 
\begin{equation}
	J_\ell = \frac{1}{K} \sum_{k \in K} 
	\big| \hat{S}_\ell(k) - S_\ell(k) \big|^2
\end{equation}
in the frequency domain is applied on the network or masking outputs $\hat{S}_\ell(k)$ during training.
The training is performed using the Adam optimizer~\cite{adam_optimizer} with standard parameters, a batch size of~$16$ and sequence length of~$50$. The initial learning rate is set to $5 \cdot 10^{-5}$ and is multiplied by~$0.6$ if the loss did not improve for~$3$ epochs. Training is stopped when the learning rate drops below~$5 \cdot 10^{-7}$ or if the loss did not improve for $10$~epochs. The network has about~$5.2$\,M parameters with only slight variations depending on the number of inputs (see next section for more details).

\section{Experimental Design}
\label{sec:variants}

While delivering higher than state of the art performance for RES and NR, a main contribution of this paper shall be a solid investigation of DNN input and output choices, thereby giving insights that can also be transferred to other approaches and network topologies.

In Figure~\ref{fig:network} four dashed signal paths can be seen at the input of the network. 
The first two are the signals that are typical inputs for a normal AEC (first stage): the microphone signal~$Y_\ell(k)$ and the reference signal~$X_\ell(k)$. The AEC then provides two further options: an estimated echo signal~$\hat{D}_\ell(k)$ and of course its enhanced signal~$E_\ell(k)$. This opens the question which input signal or combination of input signals can be used best for RES and NR with an FCRN. In~\cite{carbajal_RES_ICASSP} a similar idea was followed, however, only for a very small subset of the possible combinations, a rather simple feed-forward neural network, and without explicit consideration of noise reduction capabilities. Furthermore, an intuitive and often seen input signal combination of a RES/NR network is the pair of enhanced signal~$E_\ell(k)$ and reference signal~$X_\ell(k)$, which delivers state of the art performance, e.g., in~\cite{Halimeh_RES_ICASSP, amazon_PercepNet}. But as the reference signal does not contain any information about the room characteristics yet, could the estimated echo signal~$\hat{D}_\ell(k)$ be a better choice? Or could the microphone signal~$Y_\ell(k)$ help to improve the noise reduction as it is still unmodified in comparison to the enhanced signal~$E_\ell(k)$?
With the enhanced signal~$E_\ell(k)$ naturally being the most important input for this task, we will investigate all possible combinations with the three further options, interested in which are best to use for the FCRN as RES and NR network and where possible trade-offs can be found. 

The second major question to investigate for this application is whether it is more beneficial to perform a direct estimation of the complex output target with the network (regression) or if the nice improvements of the magnitude-bounded complex mask-based procedure in~\cite{strake_DNSchallenge_INTERSPEECH, Seidel_ySquare_arxiv} can be transferred to this task as well.

The two options are indicated by the switch right after the FCRN output layer in Figure~\ref{fig:network}. With switch position~\textit{\textbf{OutE}} a direct estimation of the real and imaginary part of the clean speech target is performed. On the other hand, with switch position~\textit{\textbf{OutM}} the task of the network is to estimate a complex mask~$M_\ell(k)$, which is then subject to amplitude compression and multiplied with the enhanced signal~$E_\ell(k)$ to obtain the clean speech target as
\begin{equation}
	\label{eq:mask}
	\hat{S}_\ell(k) =  E_\ell(k) \cdot \textrm{tanh}(|M_\ell(k)|)  \cdot \frac{M_\ell(k)}{|M_\ell(k)|}.
\end{equation}

\section{Results and Discussion}
\label{sec:results}

\begin{table*}[t]
	\centering
	\setlength\tabcolsep{0pt}
	\begin{tabular}{p{5mm}<{\centering} | p{5mm}<{\centering} | p{5mm}<{\centering}| p{5mm}<{\centering} | p{55mm}<{\centering} | p{13.5mm}<{\centering} p{13mm}<{\centering} p{13mm}<{\centering} p{13mm}<{\centering} | p{13.5mm}<{\centering} | p{13.5mm}<{\centering} | p{13.5mm}<{\centering}}
		\multicolumn{4}{c|}{\textbf{Inputs}} & & \multicolumn{4}{c|}{full mixture} & $d(n)$ & $n(n)$ & $s(n)$ \\
		$\mathbf{Y}$ & $\mathbf{X}$ & $\mathbf{\hat{D}}$ & $\mathbf{E}$ & \textbf{Model} & \textbf{PESQ} & \textbf{ERLE}$_{\text{BB}}$ & \textbf{$\Delta$SNR}$_{\text{BB}}$ & \textbf{PESQ}$_{\text{BB}}$ & \textbf{ERLE} & \textbf{$\Delta$SNR} & \textbf{PESQ} \\
		\hline
		o & o & & & Kalman (AEC) & 1.81 & 5.00 & 0.73 & 4.19 & 5.26 & --- & 4.62 \\
		\hline
		o & o & & & FCRN (AEC+RES+NR) (cf.~\cite{wang_NN_AEC_19}) & 2.26 & 18.38 & 12.61 & 3.15 & 14.35 & \textbf{20.86} & 3.52 \\
		& o & & o & Kalman (AEC) + Halimeh (RES+NR) & 2.25 & 17.78 & 12.40 & 3.30 & 16.22 & 13.24 & 2.97 \\
		\hline
		& & & o & Kalman (AEC) + FCRN (RES+NR) & 2.50 & \textbf{22.72} & \textbf{14.39} & 3.40 & \textbf{24.19} & 13.90 & 3.40 \\
		o & & & o & Kalman (AEC) + FCRN (RES+NR)  & 2.52 & \underline{21.79} & 13.38 & 3.55 & \underline{23.50} & \underline{15.33} & 3.52 \\
		& o & & o & Kalman (AEC) + FCRN (RES+NR)  & 2.46 & 19.67 & 12.75 & 3.51 & 16.78 & 12.17 & 3.45 \\
		& & o & o & Kalman (AEC) + FCRN (RES+NR)  & \underline{2.53} & 21.46 & 13.12 & \underline{3.58} & 23.45 & 12.41 & 3.62 \\
		o & o & & o & Kalman (AEC) + FCRN (RES+NR)  & \textbf{2.55} & 21.02 & \underline{13.69} & 3.51 & 21.48 & 14.77 & \underline{3.64} \\
		o &  & o & o & Kalman (AEC) + FCRN (RES+NR)  & 2.52 & 21.32 & 13.27 & \underline{3.58} & 23.44 & 13.24 & 3.53 \\
		& o & o & o & Kalman (AEC) + FCRN (RES+NR)  & 2.51 & 20.16 & 13.27 & 3.55 & 20.11 & 12.58 & 3.55 \\
		o & o & o & o & Kalman (AEC) + FCRN (RES+NR)  & 2.51 & 19.92 & 13.16 & \textbf{3.59} & 20.50 & 13.43 & \textbf{3.65} \\
	\end{tabular}
    \vspace{1mm}
	\caption{Experimental results for the different models with \textbf{direct estimation} of real and imaginary part of the \textbf{clean speech} training target~$S_\ell(k)$: ERLE and $\Delta$SNR given in~[dB], and PESQ MOS LQO. Best result per measure is in \textbf{bold} font, second best is \underline{underlined}.}
	\label{tab:results_direct}%
	\vspace{6mm}
	\centering
	\setlength\tabcolsep{0pt} 
	\begin{tabular}{p{5mm}<{\centering} | p{5mm}<{\centering} | p{5mm}<{\centering}| p{5mm}<{\centering} | p{55mm}<{\centering} | p{13.5mm}<{\centering} p{13mm}<{\centering} p{13mm}<{\centering} p{13mm}<{\centering} | p{13.5mm}<{\centering} | p{13.5mm}<{\centering} | p{13.5mm}<{\centering}}
		\multicolumn{4}{c|}{\textbf{Inputs}} & & \multicolumn{4}{c|}{full mixture} & $d(n)$ & $n(n)$ & $s(n)$ \\
		$\mathbf{Y}$ & $\mathbf{X}$ & $\mathbf{\hat{D}}$ & $\mathbf{E}$ & \textbf{Model} & \textbf{PESQ} & \textbf{ERLE}$_{\text{BB}}$ & \textbf{$\Delta$SNR}$_{\text{BB}}$ & \textbf{PESQ}$_{\text{BB}}$ & \textbf{ERLE} & \textbf{$\Delta$SNR} & \textbf{PESQ} \\
		\hline
		o & o & & & Kalman (AEC) & 1.81 & 5.00 & 0.73 & 4.19 & 5.26 & --- & 4.62 \\
		\hline
		o & o & & & FCRN (AEC+RES+NR) & 2.17 & 20.97 & 9.89 & \underline{3.31} & 16.93 & \underline{17.97} & 3.87 \\
		& o & & o & Kalman (AEC) + Halimeh (RES+NR) \cite{Halimeh_RES_ICASSP} & 2.31 & 27.23 & 14.21 & 3.20 & 25.50 & 14.24 & 3.80 \\
		\hline
		& & & o & Kalman (AEC) + FCRN (RES+NR)  & 2.37 & 32.36 & \textbf{15.28} & 3.18 & \textbf{29.16} & 14.74 & 3.84 \\
		o & & & o & Kalman (AEC) + FCRN (RES+NR) & \underline{2.52} & \underline{32.40} & 15.05 & \textbf{3.32} & 28.27 & \textbf{22.38} & 3.93 \\
		& o & & o & Kalman (AEC) + FCRN (RES+NR)  & 2.29 & 28.25 & 13.80 & 3.23 & 25.17 & 14.18 & 3.89 \\
		& & o & o & Kalman (AEC) + FCRN (RES+NR)  & \underline{2.52} & \underline{32.40} & \underline{15.27} & \underline{3.31} & 28.38 & 14.60 & 3.91 \\
		o & o & & o & Kalman (AEC) + FCRN (RES+NR)  & 2.32 & 28.84 & 13.67 & 3.21 & 26.12 & 16.66 & 3.96 \\
		o & & o & o & \textbf{Kalman (AEC) + FCRN (RES+NR)}  & \textbf{2.54} & \textbf{32.92} & \underline{15.27} & 3.30 & \underline{28.77} & 15.42 & 3.93 \\
		& o & o & o & Kalman (AEC) + FCRN (RES+NR)  & 2.48 & 30.30 & 14.61 & 3.29 & 26.73 & 14.35 & \textbf{3.98} \\
		o & o & o & o & Kalman (AEC) + FCRN (RES+NR)  & 2.49 & 30.46 & 14.77 & 3.27 & 26.78 & 14.45 & \underline{3.97} \\	
	\end{tabular}
    \vspace{1mm}
	\caption{Experimental results for the different models with \textbf{complex mask estimation} and multiplication with the enhanced signal from AEC to obtain the \textbf{clean speech} training target $S_\ell(k)$: ERLE and $\Delta$SNR given in~[dB], and PESQ MOS LQO. Only for the second row \textit{FCRN~(AEC+RES+NR)}, the estimated mask is multiplied with the \textit{microphone signal} as this experiment includes the task of AEC in the network. Best result per measure is in \textbf{bold} font, second best is \underline{underlined}. The \textbf{bold model} is \textbf{our proposal}.}
	\label{tab:results_pcRM}%
\end{table*}

We use three measures to evaluate all approaches: Speech quality in terms of wide\-band PESQ MOS LQO~\cite{ITU_P862.2, ITU_P862.2_Corr}, SNR improvement~($\Delta$SNR) in~[dB] for noise reduction, and echo suppression by echo return loss enhancement (ERLE) in~[dB] computed as in~\cite{Franzen_AEC_NetShell_ICASSP}.
To allow for a deeper understanding of the approaches, the three right-most columns provide the performance when the microphone signal consists of only one component: either echo ($d(n)$, rated with ERLE), or near-end noise ($n(n)$, rated with $\Delta$SNR), or near-end speech ($s(n)$, rated with PESQ), thereby revealing for example if a model is able to pass through a clean speech signal.
Evaluation of the normal full mixture test set is provided in the four center columns. PESQ MOS is given for the entire output signal, and---indicated by the index~\emph{BB}---the so-called black box approach according to ITU-T Recommendation~P.1110~\cite[sec. 8]{ITU_P1110} and \cite{fingscheidt_signalseparation, fingscheidt_blackbox}
is used to obtain processed \textit{components} within the output signal as $\hat{s}(n) = \tilde{d}(n) + \tilde{n}(n) + \tilde{s}(n)$. By that, an evaluation of ERLE, $\Delta$SNR, and PESQ is also possible on the separated processed components $\tilde{d}(n)$, $\tilde{n}(n)$, and $\tilde{s}(n)$ for full mixture inputs even in double talk. 

Table~\ref{tab:results_direct} provides experimental results for direct estimation of the clean speech. 
Starting in the first row, the performance of the Kalman filter as first stage is given. With about~5\,dB, its echo suppression is rather moderate (as its internally related postfilter is omitted for this work and nonlinearities are part of the simulation setup), but remains a nearly perfect near-end speech quality.

The following rows provide two approaches as baseline comparisons: First, a single-stage 'all-in-one' approach trains the FCRN to perform all tasks (AEC+RES+NR) at once, as proposed in~\cite{wang_NN_AEC_19} using direct estimation and further elaborated in~\cite{Franzen_AEC_NetShell_ICASSP}. As for the Kalman filter, the only available inputs to this approach are the microphone signal~$\mathbf{Y}$ and reference signal~$\mathbf{X}$ (for brevity denoted as vectors without frame and bin index here). It shows a good performance in terms of echo and noise reduction with up to nearly $21$\,dB, but at the cost of a degraded near-end speech ($3.52$ MOS LQO when only $s(n)$ is present at the microphone). 
The second baseline is our reimplementation of the model of Halimeh~et~al.~\cite{Halimeh_RES_ICASSP} for RES and NR. To allow for comparability it is deployed after our Kalman filter as second stage with the in- and output setup described in this work. The inputs to the network are given as reference signal~$\mathbf{X}$ and enhanced signal~$\mathbf{E}$~\cite{Halimeh_RES_ICASSP}. In addition to the Kalman filter, a good improvement in ERLE as well as $\Delta$SNR can be observed with this model. While the overall PESQ score is about the same as for the \textit{FCRN~(AEC+RES+NR)}, however, the Halimeh model is by far not able to maintain the near-end speech quality in this regression setup.

In contrast, when the FCRN is deployed as second stage, the near-end speech quality is up to $0.68$\,MOS points higher ($3.65$) and the overall PESQ score improves by up to~$0.3$ to $2.55$\,MOS points. Results for all possible inputs combined with~$\mathbf{E}$ are provided: 
While the single input~$\mathbf{E}$ gives a strong echo and noise reduction, but at a somewhat lower near-end speech quality, the other results reveal interesting patterns that can easily be seen in the columns for $d(n)$ and $n(n)$: 
\textit{First, a higher ERLE is obtained if the estimated echo~$\mathbf{\hat{D}}$ is used instead of the reference signal~$\mathbf{X}$. Second, noise reduction is always improved if the microphone signal~$\mathbf{Y}$ is used as an additional input.} Interestingly, and contrary to what one might expect, these effects diminish if simply all inputs are used. We conclude that a suitable preselection of the inputs is helpful and key to tweak the desired network behavior. 

Table~\ref{tab:results_pcRM} provides results for the same experiments but now with complex mask estimation and multiplication with the enhanced signal~$\mathbf{E}$ according to~(\ref{eq:mask}). For the second row \textit{FCRN~(AEC+RES+NR)}, however, the estimated mask is multiplied with~$\mathbf{Y}$, as this experiment includes the task of AEC in the network.
\textit{It can be seen that all approaches benefit from the masking procedure.} Also the Halimeh model now obtains an improved near-end speech quality which is in line with its original design as masking approach~\cite{Halimeh_RES_ICASSP}. 
Investigating the various input combinations of the FCRN, \textit{we see again the pattern of a higher ERLE when using~$\mathbf{\hat{D}}$ instead of~$\mathbf{X}$, and using~$\mathbf{Y}$ in addition supports a stronger	noise reduction} (except all four inputs are used).

If no specific behavior is desired but the best trade-off should be obtained, we identify the input combination of~$\mathbf{Y}$, $\mathbf{\hat{D}}$ and~$\mathbf{E}$ for the FCRN with complex masking~(\ref{eq:mask}) to give the best overall results, see the model in bold font in Table~\ref{tab:results_pcRM}.
It scores four metrics on~1st and~2nd rank, with a strong near-end speech only PESQ of more than $3.9$\,MOS points. 
With an improved ERLE and $\Delta$SNR, and a $0.23$ higher full mixture PESQ score than Halimeh ($2.54$\,PESQ points), this is in line with informal subjective listening experience. 

\section{Conclusions}
\label{sec:conc}
\vspace{1mm}

We presented a hybrid speech enhancement scheme with a novel multi-input fully convolutional recurrent network (FCRN) for deep residual echo suppression and noise reduction. In an extensive ablation study on network input signal combinations, we surprisingly found that the often used loudspeaker reference signal should not be used for this task, but the enhanced signal after acoustic echo cancellation (AEC), the microphone signal, and the echo estimate. We improve previous approaches by more than $0.2$\,PESQ points and provide insights to simple design choices, such as direct estimation or complex mask estimation.

\bibliographystyle{IEEEbib}
\bibliography{biblio_franzen}

\end{document}